\documentclass{ws-procs9x6}

\begin{document}

\title{SEARCH FOR A POSSIBLE SPONTANEOUS EMISSION OF MUONS FROM HEAVY NUCLEI}

\author{M. GIORGINI$^*$} 

\address{University of Bologna and INFN Bologna \\
Viale B. Pichat 6/2, I-40127 Bologna, Italy\\
$^*$E-mail: miriam.giorgini@bo.infn.it \\ ~\\
Talk given at the $11^{th}$ ICATPP Conference on
Astroparticle, Particle, Space Physics, Detectors and Medical Physics 
Applications, Como (Italy), 5-9 October 2009.}

\begin{abstract}
A search for an exotic natural radioactivity of lead 
nuclei, using nuclear emulsion sheets as detector, is described. We 
discuss the 
experimental set-up of a test performed at the Gran Sasso National 
Laboratory (Italy), the event simulation, data analysis and  
preliminary results.
\end{abstract}


\bodymatter

\section{Introduction}\label{sec:intro}

In the late 80's, some theoretical work was dedicated to investigate possible
exotic types of nuclear radioactivity, consisting in the emission of
light particles such
as pions or muons from heavy nuclei \cite{ion1,ion2}.

 Muons or pions could be emitted by nuclei through the decays\cite{ion2}:
\begin{equation}
(A,Z) \rightarrow \mu^\pm + \nu_\mu(\overline \nu_\mu)+(A_1,Z_1)+...+(A_n,Z_n)
\label{eq:1}
\end{equation}
\vspace{-7mm}
\begin{equation}
(A,Z) \rightarrow \pi^\pm (\pi^0)+(A_1,Z_1)+...+(A_n,Z_n)
\label{eq:2}
\end{equation}
where, for reasons of energy and momentum conservation, the number
of fragments $n$ is $\geq 2$. The nuclei $(A_1,Z_1),~(A_2,Z_2),...,(A_n,Z_n)$ 
would yield a sequence of $\beta^-$ decays leading finally to stable nuclei 
with a balanced neutron-proton ratio.

In ref. 2 some nuclear charge thresholds 
for different possible spontaneous particle emission were listed: \par
 $(i)~ \mu^\pm$ (prompt muon) for $Z \geq 72$ \par
 $(ii)~ \pi^\pm$ (prompt pion $\rightarrow$ delayed muon) for $Z \geq 76$ \par 
 $(iii)~ 2\mu^\pm$ (prompt muon pairs) for $Z \geq 91$ \par
 $(iv)~2\pi^\pm$ (prompt pion pairs $\rightarrow$  delayed muon pairs) 
for $Z \geq 100$ 

Natural lead is mainly composed by three nuclides: $^{206}$Pb (24.1\%), 
 $^{207}$Pb (22.1\%) and $^{208}$Pb (52.4\%). They are stable nuclides, but 
the channels $(i)$ and $(ii)$ are energetically allowed. 

In the hypothesis of the decays (\ref{eq:1}) and (\ref{eq:2}), the 
fission fragments would remain nearly at rest; most of 
the available energy would be used to produce the $\mu$ (or $\pi$) and
the kinetic energies of $\mu$ and $\nu_\mu$ (or $\pi$).
The total kinetic energy $Q_\mu$ in a decay in 2 fragments
with close values of $A_1$ and $A_2$ (symmetric fission) is $\sim$30 MeV 
for negative muons and $\sim$20 MeV for positive muons. 
 Considering that the associated muon neutrino takes away a sizable fraction
of this energy, the spectrum of emitted muons could be like in a 
$\beta$ decay, with an average energy of $10 \div 15$ MeV. 

The spontaneous or neutron induced fission of Pb has never been observed.
A search for a Pb muonic decay can be made as a byproduct of the OPERA 
experiment\cite{opera}, aimed to confirm neutrino oscillations in the 
parameter region indicated by some atmospheric neutrino experiments 
\cite{soudan}-\cite{superk}. In the following, we discuss the 
experimental set-up, a Monte Carlo (MC) 
simulation and the reachable limits.

\vspace{-6mm}

\section{Experimental set-up}
\label{sec:setup}
We propose to perform an experimental search for muonic radioactivity from
lead nuclei in the low background conditions offered by the Gran Sasso 
underground Laboratory (LNGS). The low cosmic muon flux and the 
low natural radioactivity of the rock in the experimental 
halls of the LNGS provide unique conditions, allowing a potential
discovery, or, at least, to establish a good upper limit for
this exotic decay process.
 A detailed description of the different background sources is given in 
ref. 7.

The OPERA base element (``brick'') is composed of 56 lead sheets (1 mm thick) 
interleaved with 2 nuclear emulsion sheets (43 $\mu$m thick) on both sides 
of a 200 $\mu$m thick plastic base. 
The area of each sheet is $10.3 \times 12.8$ cm$^2$.
The OPERA bricks (each containing a mass of 8.23 kg of Pb) could 
allow an experimental search for 
muon emission from lead, with exposures of several months. Their
analyses with the fast automated optical microscopes\cite{microscopi} 
would establish the local background contributions and validate the 
analysis procedures.

 As the background rejection/reduction (see ref. 7) 
is a crucial point for this search, the detectors are surrounded on 
all sides by a shield, making a closed box structure. The shield 
is composed of an inner layer 5 cm thick of 
very pure copper followed by 15 cm of very low 
activity lead. 
 The third layer of the shield is a 3 cm thick polyethylene, in order to 
absorb neutrons\footnote{These thicknesses have been found to be adequate with 
a Monte Carlo simulation considering the effects of $0.5 \div 2.6$ 
MeV photons, which  could produce electrons mimicking the searched events.}.
 The set-up is located in the emulsion storage room, in hall B of the Gran 
Sasso Laboratory. The radon reduction is obtained with a ventilation 
system with fresh air forced circulation.

\vspace{-6mm}
\section{Monte Carlo simulation}
\label{sec:MC}

A MC simulation program was implemented to estimate the occurrence of 
different event topologies. The simulation is based on the GEANT \cite{geant}
 package applied to the OPERA lead/emulsion set-up.
The simulation reproduces one complete OPERA brick, where muons of different 
energies (see the first column of Table \ref{tab:mu}) 
originate in random positions in the lead sheets. The initial muon 
directions are isotropically generated. We assumed different definitions 
for a candidate event, requiring that the muon crosses at least: $(i)$ 2 
emulsion films, $(ii)$ 4 emulsion film
(and thus also a lead plate) and $(iii)$ 6 emulsion films (and thus also 
2 lead plates). We also requested the 
detection of the decay positron or electron in at least $(iv)$ 2, $(v)$ 
4 or $(vi)$ 6 emulsion layers, together with the muon detection.

\begin{figure}[ht]
\vspace{-5mm}
\begin{center}
       \mbox{\epsfysize=3.5cm
             \epsffile{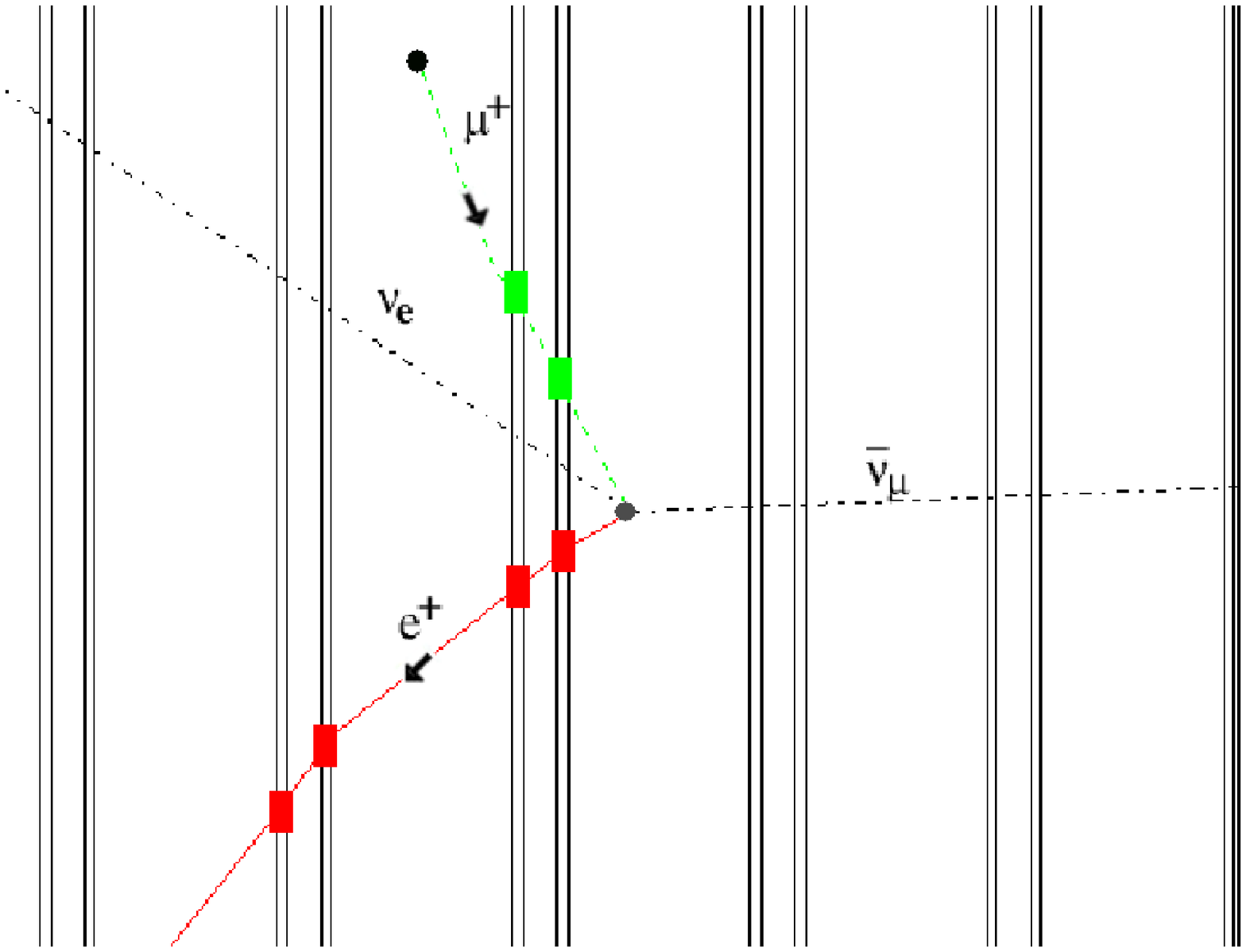}
               \hspace{0.7cm}
               \epsfysize=4.0cm
               \epsffile{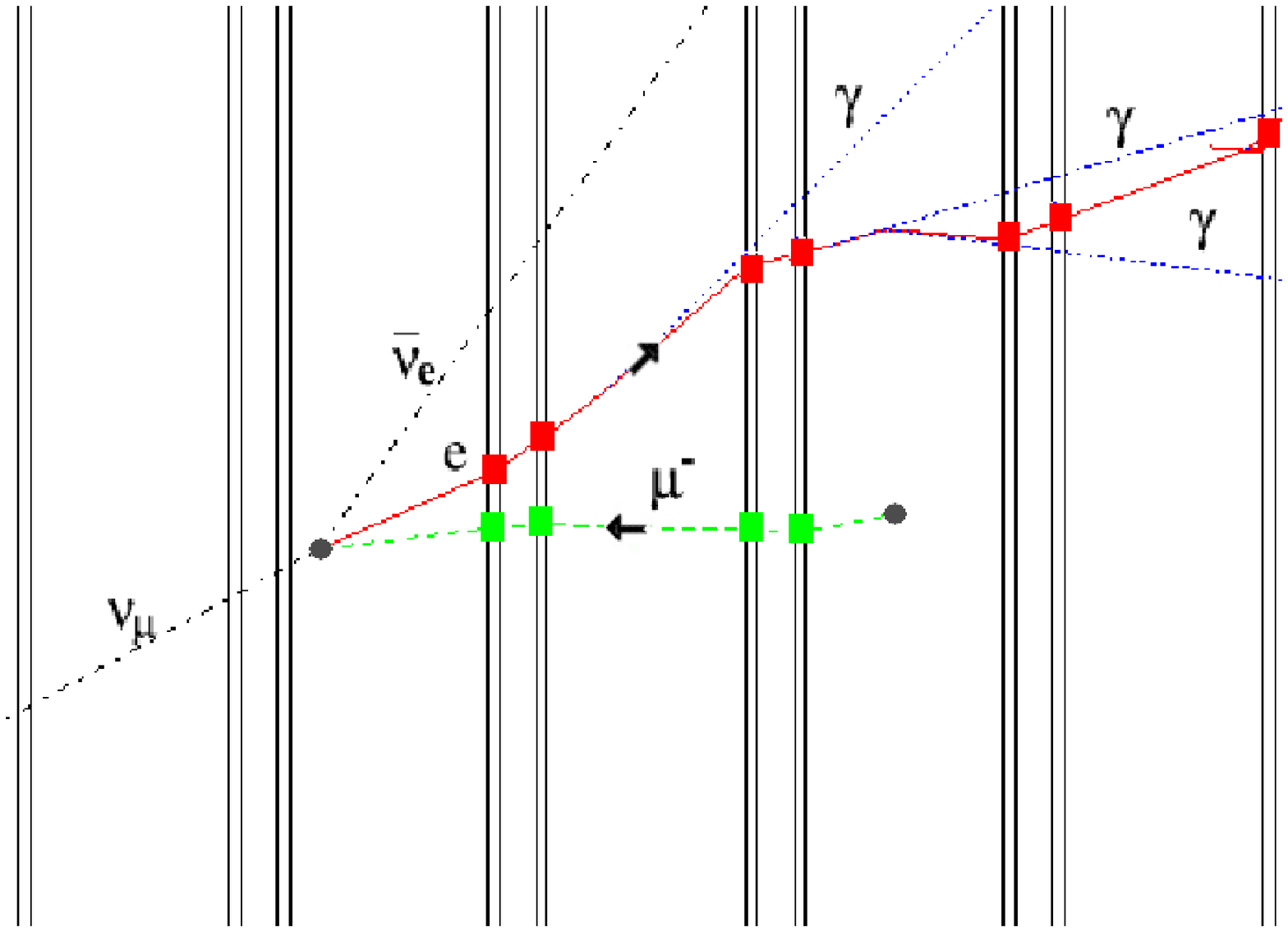}}
\caption{Simulated events of spontaneous $\mu^+$ (left) and $\mu^-$ (right)
emission from lead inside an OPERA brick, assuming an initial muon kinetic 
energy of 15 MeV.}
\label{fig:mu}
\end{center}
\end{figure}

\vspace{-5mm}
For each topology, the values quoted 
in Table \ref{tab:mu} were obtained as averages over all
 possible muon emission directions. These estimates may be considered 
as geometrical efficiencies $\epsilon_g$.  The percentages of events listed in 
Table \ref{tab:mu} were computed for samples of 1500 MC events 
with fixed energies.

In Fig. \ref{fig:mu} are shown two simulated 
muon emission events from the lead inside an OPERA brick, assuming an 
initial kinetic energy of 15 MeV, and the production of a $\mu^+$ (left) 
and a $\mu^-$ (right).

\begin{table}
\tbl{Percentage of events as a function 
of the minimum number of emulsion films crossed by both $\mu$ $(N_\mu)$ 
and $e$ $(N_e)$, for different $\mu$ initial energies (col. 1).}
{\begin{tabular}{|c|c|c|c|c|c|c|c|c|c|c|c|c|}\Hline
$N_{\mu}$ & 2 & 2 & 2 & 2 & 4 & 4 & 4 & 4 & 6 & 6 & 6 & 6 \\
$N_{e}$   & 0 & 2 & 4 & 6 & 0 & 2 & 4 & 6 & 0 & 2 & 4 & 6 \\\hline
5 MeV  & 12 & 10 & 9  & 7  & - & - & - & - & - & - & - & - \\
7 MeV  & 25 & 23 & 21 & 14 & - & - & - & - & - & - & - & -  \\
10 MeV & 51 & 46 & 41 & 31 & 0.6 & 0.5 & 0.4 & 0.3 & - & - & - & - \\
15 MeV & 74 & 66 & 58 & 44 & 31 & 28 & 23 & 17 & 1 & 1 & 0.9 & 0.7  \\
20 MeV & 84 & 72 & 63 & 49 & 56 & 50 & 40 & 30 & 29 & 26 & 21 & 15  \\
25 MeV & 89 & 75 & 66 & 50 & 70 & 60 & 49 & 36 & 48 & 42 & 34 & 25 \\
30 MeV & 92 & 75 & 67 & 52 & 78 & 66 & 54 & 36 & 60 & 51 & 42 & 28 \\
\Hline
\end{tabular}
}
\label{tab:mu}
\end{table}

\section{Estimates of global detection efficiencies}
\label{sec:eff}
 The geometrical $\epsilon_g$ have to be multiplied by the $\mu^\pm$ and 
$e^\pm$ reconstruction efficiencies to obtain the total efficiency.
 With the OPERA tracking procedure \cite{saverio}, the mean 
detection efficiency for each ``microtrack'' in one emulsion 
film is $\simeq 95\%$.  The instrumental limit of 0.8 rad on the incident 
direction introduces an event selection of $\sim$30\%.
The ``base track'' (obtained from 2 microtracks separated by 
the plastic base) reconstruction efficiency is 
$\simeq 90\%~(95\% \times 95\%)$. The 
top-bottom linking efficiency, mainly due to the multiple scattering in the 
plastic base, ranges from the $\sim$50\% 
for 5 MeV particles to $\sim$99\% for particles with energies $\geq 20$ MeV.
 The base track linking efficiency is $\sim$6\% for the whole range 
of initial muon energies. 

 The detection efficiency $\epsilon_{tot}$ 
for a muon crossing at least $n$ emulsion 
films is the product of the percentage of events with these topological 
requirements $\epsilon_g$ (col. 6 of Table \ref{tab:mu}), the 30\% 
given by the angular limit, $n$ times the 95\% 
efficiency for the microtracks, $n/2$ the top-bottom linking efficiency and
the linking efficiency between the base tracks. 
Table \ref{tab:eff} shows the \% efficiencies 
$\epsilon_g$ and $\epsilon_{tot}$ computed requiring
at least 2 emulsion films crossed by the muon (col. 2-3) and 4   
emulsion films crossed by the muon (col. 4-5). The values refer to the present
OPERA tracking \cite{saverio}.

As the expected halflifes $t_{1/2}$ are much larger 
than any reasonable exposure time $T$, the expected sensitivities are 
estimated from
\begin{equation}
\frac{\delta N}{N_0} = \frac{\ln 2}{t_{1/2}}T \epsilon_{tot}
\label{eq:6}
\end{equation}
where $\delta N = 2.3$ is the number of events corresponding to a 90\% C.L. 
limit assuming no candidates, $N_0$ is the initial number of nuclei, and 
$\epsilon_{tot}$ is the experimental efficiency. 
Using one OPERA brick for one year exposure and a global 
detection efficiency of $\sim$10\%, a 
sensitivity of $\sim$$7 \cdot 10^{23}$ yr ($90\%$ C.L.) could be reached.

\begin{table}
\tbl{Geometric efficiencies $(\epsilon_g)$ and total muon detection 
efficiencies $(\epsilon_{tot})$ as a function of  
the initial muon energy computed on the basis of the OPERA
tracking procedure for $N_\mu=2$ (col. 2-3) and for  
$N_\mu=4$ (col. 4-5). All values are in \%.}
{\begin{tabular}{|c||c|c||c|c|}\Hline
&\multicolumn{2}{c||}{$N_\mu=2$}&\multicolumn{2}{c|}{$N_\mu=4$} \\
\hline
$E_\mu$ & $\epsilon_g$ & $\epsilon_{tot}$ & 
$\epsilon_g$ & $\epsilon_{tot}$  \\
\hline
\hspace{0.5cm} 5 MeV \hspace{0.5cm} & \hspace{0.5cm} 12 \hspace{0.5cm} 
& \hspace{0.5cm} 1.5 \hspace{0.5cm}  & \hspace{0.5cm} - 
\hspace{0.5cm} & \hspace{0.5cm} - \hspace{0.5cm}   \\
7 MeV  & 25  & 4.3   &  -  & -         \\
10 MeV & 51  & 12    & 0.6 & 0.006     \\
15 MeV & 74  & 19    & 31  & 0.4       \\
20 MeV & 84  & 22    & 56  & 0.8        \\
25 MeV & 89  & 24    & 70  & 1          \\
30 MeV & 92  & 25    & 78  & 1.1        \\
\Hline
\end{tabular}}
\label{tab:eff}
\end{table}

\section{Conclusions and perspectives}
\label{sec:conclu}
A test search for spontaneous emission of muons from 
Pb nuclei, using some OPERA lead/emulsion bricks, was described.
We are in the process of making a complete simulation of the detector 
including its response and the track reconstruction efficiencies. We 
have shown that stringent limits for spontaneous muon radioactivity
may be reached: $t_{1/2} \geq  7 \cdot 10^{23}$ years.

\vspace{3mm}
\noindent {\small {\bf Acknowledgments.} I would like to acknowledge 
the cooperation of all the members of the Bologna 
group, in particular of Dr. V. Togo.}

\bibliographystyle{ws-procs9x6}
\bibliography{ws-pro-sample}

\vspace{-4mm}
\begin{thebibliography}{5}

\bibitem{ion1} D.B. Ion, R. Ion-Mihai and M. Iva\c{s}cu, 
%{\em Spontaneous pion emission as a new natural radioactivity}, 
{\em Ann. Phys.} {\bf 171}, 237 (1986).%-252.

\bibitem{ion2} D.B. Ion, R. Ion-Mihai and M. Iva\c{s}cu, 
%{\em Spontaneous muon emission as new natural radioactivity}, 
{\em Rev. Roum. Phys.} {\bf 31}, 209 (1986).%-212.

\bibitem{opera} {\it http://operaweb.lngs.infn.it/.} \\
R. Acquafredda et al., {\em New Journal of Physics} {\bf 8}, 303 (2006);
%{\it First events from the CNGS neutrino beam detected
%in the OPERA experiment} 
{\em Journal of Instr.} {\bf 4}, P04018 (2009).
%{\it The OPERA experiment in the CERN to Gran Sasso neutrino beam} 

\bibitem{soudan} M.C. Sanchez et al., {\em Phys. Rev.} {\bf D68}, 
 113004 (2003). \\
W.W.M. Allison et al., {\em Phys. Rev.} {\bf D72}, 052005 (2005).

\bibitem{macro}
M. Ambrosio et al., {\em Phys. Lett.} {\bf B434}, 451 (1998); 
{\em Phys. Lett.} {\bf B478}, 5 (2000);
{\em Phys. Lett.} {\bf B517}, 59 (2001);
{\em Eur. Phys. J.} {\bf C36}, 323 (2004).

\bibitem{superk} J. Hosaka et al., {\em Phys. Rev.} {\bf D74}, 032002 
(2006). \\
K. Abe et al., {\em Phys. Rev. Lett.} {\bf 97}, 171801 (2006).

\bibitem{exod} L. Arrabito et al., {\em hep-ex/0506078} (2005).

\bibitem{microscopi} S. Aoki et al., {\em Nucl. Instr. Meth.} {\bf A473}, 
 192 (2001). \\ 
N. D'Ambrosio et al., {\em Nucl. Instr. Meth.} {\bf A525}, 193 (2004).

\bibitem{geant}% R. Brun et al., GEANT, CERN report DD/EE84-1 (1987).
{\it http://wwwasd.web.cern.ch/wwwasd/geant/}.

\bibitem{saverio} N. Armenise et al.,  {\em Nucl. Instr. Meth.} {\bf A551}, 
261 (2005). \\  
%{\it High-speed particle tracking in nuclear emulsion by last-generation 
%automatic microscopes}
L. Arrabito et al., {\em Nucl. Instr. Meth.} {\bf A568}, 578 (2006);
%{\it Hardware performance of a scanning system for high speed analysis 
%of nuclear emulsions} 
 {\em Journal of Instr.} {\bf 2}, P05004 (2007). \\
%{\it Track reconstruction in the emulsion-lead target of the OPERA 
%experiment using the ESS microscope} 
N. Agafonova et al., {\em Journal of Instr.} {\bf 4}, P06020 (2009). \\
%{\it The detection of neutrino interactions in the 
%emulsion/lead target of the OPERA experiment}

\end{thebibliography}

\end{document}